\documentclass[12pt]{article}
\usepackage{epsf}
\usepackage{graphicx}
\usepackage{a4}
\usepackage{amsmath}
\usepackage{amssymb}
\usepackage{cite}		
\usepackage{color}

\def\hybrid{\topmargin 0pt
        \oddsidemargin 0pt
        \headheight 0pt \headsep 0pt
        \textwidth 6.25in       
        \textheight 9.5in       
        \marginparwidth .875in
        \parskip 5pt plus 1pt   \jot = 1.5ex}

\catcode`\@=11
\def\marginnote#1{}

\newcount\hour
\newcount\minute
\newtoks\amorpm
\hour=\time\divide\hour by60
\minute=\time{\multiply\hour by60 \global\advance\minute by-\hour}
\edef\standardtime{{\ifnum\hour<12 \global\amorpm={am}%
        \else\global\amorpm={pm}\advance\hour by-12 \fi
        \ifnum\hour=0 \hour=12 \fi
        \number\hour:\ifnum\minute<10 0\fi\number\minute\the\amorpm}}
\edef\militarytime{\number\hour:\ifnum\minute<10 0\fi\number\minute}

\def\draftlabel#1{{\@bsphack\if@filesw {\let\thepage\relax
   \xdef\@gtempa{\write\@auxout{\string
      \newlabel{#1}{{\@currentlabel}{\thepage}}}}}\@gtempa
   \if@nobreak \ifvmode\nobreak\fi\fi\fi\@esphack}
        \gdef\@eqnlabel{#1}}
\def\@eqnlabel{}
\def\@vacuum{}
\def\draftmarginnote#1{\marginpar{\raggedright\scriptsize\tt#1}}

\def\draft{\oddsidemargin -.5truein
        \def\@oddfoot{\sl preliminary draft \hfil
        \rm\thepage\hfil\sl\today\quad\militarytime}
        \let\@evenfoot\@oddfoot \overfullrule 3pt
        \let\label=\draftlabel
        \let\marginnote=\draftmarginnote
   \def\@eqnnum{(\theequation)\rlap{\kern\marginparsep\tt\@eqnlabel}%
\global\let\@eqnlabel\@vacuum}  }

\def\numberbysection{\@addtoreset{equation}{section}
        \def\theequation{\thesection.\arabic{equation}}}

\def\titlepage{\@restonecolfalse\if@twocolumn\@restonecoltrue\onecolumn
     \else \newpage \fi \thispagestyle{empty}\c@page\z@
        \def\thefootnote{\fnsymbol{footnote}}
	\setcounter{page}{0} }
\def\endtitlepage{\if@restonecol\twocolumn \else  \fi
        \def\thefootnote{\arabic{footnote}}
        \setcounter{footnote}{0}}  
\definecolor{c1}{rgb}{1, 0, 0}
\definecolor{c2}{rgb}{0, 1, 0}
\definecolor{c3}{rgb}{0, 0, 1}
\definecolor{c4}{rgb}{1, 0, 1}
\definecolor{c5}{rgb}{0, 1, 1}

\catcode`@=12
\relax

\def\beq{\begin{equation}}
\def\eeq{\end{equation}}
\def\bea{\begin{eqnarray}}
\def\eea{\end{eqnarray}}
\def\EQ{\begin{equation}}
\def\EN{\end{equation}}

\relax
\numberbysection
\hybrid

\begin{document}
\begin{center}
{\large\bf Spin clusters and conformal field theory}\\[.3in] 

{\bf G. Delfino$^1$, M.\ Picco$^{2}$, R.\ Santachiara$^{3}$ and J. Viti$^4$}\\
$^1$ \it{SISSA and INFN,\\
                       via Bonomea 265,\\
                       34136 Trieste, Italy,\\
              e-mail: {\tt delfino@sissa.it}. }\\
$^2$ {\it LPTHE\/}\footnote[5]{Unit\'e mixte de recherche du CNRS 
UMR 7589}, 
        {\it  Universit\'e Pierre et Marie Curie-Paris6\\
              Bo\^{\i}te 126, Tour 13-14, 5 \`eme \'etage, \\
              4 place Jussieu,
              F-75252 Paris CEDEX 05, France, \\
    e-mail: {\tt picco@lpthe.jussieu.fr}. }\\
  $^3$ {\it LPTMS\/}\footnote[6]{Unit\'e mixte de 
             recherche du CNRS UMR 8626}, 
             \it{Universit\'e Paris-Sud\\
             B\^atiment 100\\
          F-91405 Orsay, France, \\
    e-mail: {\tt raoul.santachiara@lptms.u-psud.fr}. }\\
   $^4${\it LPT} 
   \it{Ecole Normale Superieure \& CNRS,\\
             24 Rue Lhomond,\\
          Paris CEDEX 05, France\\
    e-mail: {\tt jacopo.viti@lpt.ens.fr}. }\

\end{center}
\centerline{(Dated: \today)}
\vskip .2in
\centerline{\bf ABSTRACT}
\begin{quotation}
\noindent
We study numerically the fractal dimensions and the bulk three-point connectivity for the spin clusters
of the $Q$-state Potts model in two dimensions with $1\leq Q~\leq~4$.
We check that the usually invoked correspondence between FK clusters and spin clusters works at the level of fractal dimensions. However, the fine structure of
the conformal field theories  describing critical clusters first manifests at the level of the
three-point connectivities.  
Contrary to what recently found for FK clusters, no obvious relation emerges for generic $Q$ between
the spin cluster connectivity and the structure constants obtained from analytic continuation of the minimal model ones.
The numerical results strongly suggest then that spin and FK clusters are described by conformal field theories with different
realizations of the color symmetry of the Potts model.

\vskip 0.5cm 
\end{quotation}

\newpage
\section{Introduction}
Percolation theory deals with clusters and their connectivity properties \cite{SA}.
It represents a chapter of the theory of critical phenomena because, changing a control parameter, 
it is possible to find a critical value beyond which the probability that a point belongs to an infinite cluster 
is non-zero. Fundamental quantities of the theory are the connectivity functions $P_n(x_1,\ldots,x_n)$, giving the
probabilities that $n$ points with coordinates $x_1,\ldots,x_n$ belong to the same cluster. If at the critical point
the connectivities decay algebraically for large separations among the points, the percolation transition is continuous and
is expected to exhibit universal properties described by a local field theory. 
Within such a theory the connectivities are related to the $n$-point functions of some field $\phi(x)$ whose scaling dimension
$X$ determines the two-point connectivity
\EQ
P_2(x_1,x_2)\propto |x_1-x_2|^{-2X}
\label{P2}
\EN
up to a non-universal normalization; within the usual lattice formulation of percolation problems, this relation is
intended for distances between the points much larger than the lattice spacing. The fractal dimensions of the clusters
is $d-X$ in $d$ dimensions. On the other hand, in a local isotropic field theory scale invariance is upgraded to conformal
invariance \cite{IZ}, and the simplest implication of conformal field theory (CFT) is that the critical three-point connectivity
takes the form \cite{Polyakov}
\EQ
P_3(x_1,x_2,x_3)=R\,\sqrt{P_2(x_1,x_2)P_2(x_1,x_3)P_2(x_2,x_3)}\,,
\label{P3}
\EN
with $R$ an universal connectivity constant. The scaling dimensions and the structure constants like $R$ are fundamental data
for solving a  CFT. 

No exact result is available for universal percolative properties in three dimensions. Even in two dimensions,
where the conformal symmetry group is infinite dimensional \cite{BPZ}, the situation is theoretically unsatisfactory.
The reason is that percolative critical points 
are not described by those CFT minimal models \cite{BPZ}  where the realization of conformal symmetry is fully understood.
The best illustration comes from the clusters of the $Q$-state Potts model, which have been the most studied and are also the
subject of this paper. For $Q<4$, the $Q$-state Potts model, defined below in \eqref{hamiltonian},  undergoes  a continuous
magnetic phase transition. 

It is known since the renormalization group analysis of \cite{CK} that the two-dimensional Ising model
($Q=2$) admits two different universality classes of percolative behavior, describing percolative clusters with
different fractal dimensions. One universality class is represented by the natural clusters
(referred to as geometrical clusters, or simply spin clusters), obtained connecting the nearest neighboring sites with
the same color. The other describes  the so-called Fortuin-Kasteleyn (FK) clusters \cite{FK} (sometimes referred to as droplets),
obtained connecting nearest neighboring sites with the same color with 
probability $1-e^{-J}$, $J>0$ being the Potts coupling. Both the spin and FK clusters become critical
at the magnetic transition point\footnote{
Notice that this seems to be peculiar to the two-dimensional Potts model.  
For other two-dimensional models spin and FK clusters are not necessarily critical 
at the same temperature, see for instance \cite{PSS}. In three dimensions Ising spin clusters percolate above
$J_c$, as first observed numerically in \cite{M-K} (see also \cite{isingperc} and references therein).} $J_c$ 
and are described by
CFTs with the same central charge.
A more general study \cite{CP} then showed
that FK clusters remain critical at $J_c$ for real values of $Q$ (in the limit $Q\rightarrow 1$ they correspond to the
clusters of random percolation \cite{FK}) and pointed to the existence of a similar critical line for spin clusters. 

The issue of the determination of the constant $R$ in (\ref{P3}) for FK clusters was considered in \cite{DV}.
It was argued there that\footnote{Throughout the paper the subscripts $FK$ and $S$ refer to FK and spin clusters, respectively.} 
\EQ
R_{FK}=\sqrt{2}\,C(X_{FK})~.
\label{Rfk}
\EN
We denote by $C(Y)$ the structure constant for three fields with scaling dimension $Y$, obtained in the most general CFT
with a non-degenerate  spectrum  (i.e. no additional symmetries) in the central charge range $c\leq 1$ of interest for the Potts model. 
The $S_Q$ (permutational) symmetry of the Potts model was argued to produce the prefactor $\sqrt{2}$.
The  structure constants for the above mentioned CFT with non-degenerate spectrum first appeared in \cite{AlZ,KoPe} and have also been recently
rederived in \cite{PSVD}, where through Coulomb gas techniques it was shown how they can be obtained from analytic
continuation of the minimal model ones \cite{DF85}. The prediction \eqref{Rfk} has been confirmed numerically
in \cite{ZSK} for $Q=1$ (random percolation)
and in \cite{PSVD} for generic (i.e. also non-integer) values of $Q$.

The relevance of the connectivity constant $R$ from the point of view of CFT is easily illustrated through the example
of $Q=2$. In the Ising model, $X_{FK}=1/8$ and coincides with the dimension of the spin field,  belonging
to the conformal spectrum of the $c=1/2$ minimal conformal model.
This minimal model, which describes the Ising magnetic phase transition,
contains no information about $R_{FK}$: indeed the structure constant of three spin fields $C_{min}(1/8)$ vanishes,
$C_{min}(1/8)=0$, consistently with $Z_2$ invariance of the lattice Hamiltonian. On the other hand $C(1/8)$ takes a non-zero
value that can be traced back to the fact that correlation functions in a non-minimal CFT are not always required to satisfy the same
differential equations of a minimal model CFT, \cite{PSVD}. The value of $C(1/8)$ accounts for the three-point FK connectivity and shows that
percolative properties are described by a non-minimal realization of conformal symmetry with that value of central charge.

In view of the present limited understanding of non-minimal $c<1$ conformal field theories,
the identification of percolation phenomena as physical realizations and the possibility of performing numerical studies
play a significant theoretical role. In this paper we investigate numerically (\ref{P3}) for the Potts spin clusters,
which are theoretically less understood than FK clusters. Indeed, as we recall in the next section,
the fractal dimension of spin clusters is only conjectured for $Q\neq 2$. A picture which includes this conjecture
and is often adopted in the literature sees the spin clusters as equivalent to the FK clusters of the tricritical
branch of the Potts model. Critical and tricritical branches are known to meet at $Q=4$ and can be seen as continuation
of each other, as we will discuss in the next section.
The values $\sqrt{2}\,C(X_S)$ for $Q=2,3,4$ were then quoted in \cite{DV} for comparison with future numerical determination of $R_S$. Here we evaluate $R_S$ by Monte Carlo simulations for several values of $Q$ in the range $1\leq Q \leq 4$
and show that $R_S\neq\sqrt{2}\,C(X_S)$ for $Q<4$,
meaning that a simple continuation of FK results, usually adopted at the level of fractal dimensions,
fails for the connectivity constant $R$.
Deferring the discussion of our findings to the last section, we give additional details 
on the theoretical background in the next section and present the numerical results in section~3.

\section{Potts clusters}
\begin{figure}[t]
\begin{center}
\epsfxsize=300pt\epsfysize=350pt{\epsffile{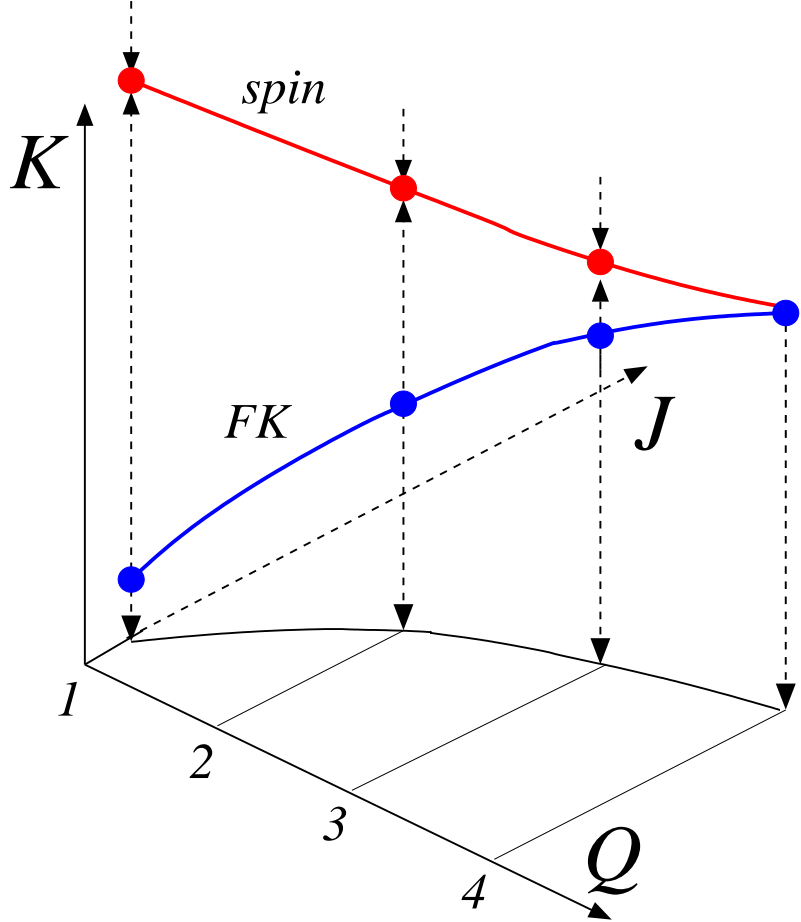}}
\end{center}
\caption{Schematic phase diagram for generalized clusters in
the $Q$-state Potts model. Vertical arrows indicate renormalization group flows towards the spin cluster fixed line and away from the FK fixed line.
}
\label{PD}
\end{figure}

The $Q$-state Potts model is defined by the lattice Hamiltonian \cite{Potts,Wu}
\EQ
{\cal H}_{Potts}=-J\sum_{\langle x,y\rangle}\delta_{s(x),s(y)}\,,\hspace{1cm}s(x)=1,\ldots,Q\,
\label{hamiltonian}
\EN
and for a critical value $J_c$ of the coupling exhibits a ferromagnetic phase transition
which is of second order for $Q\leq 4$ \cite{Baxter}. We will refer to the $Q$ values of the spin variable $s(x)$ as
$Q$ different colors. It is useful to define generalized clusters obtained connecting nearest neighboring
sites with the same color with probability $p=1-e^{-K}$, $K>0$ being a parameter. In  \cite{CP}  a description of
these clusters was given in terms of  the Hamiltonian
\beq
\label{hamDP}
\mathcal{H}=\mathcal{H}_{Potts}-K\sum_{\langle x,y\rangle}\delta_{s(x),s(y)}\delta_{\tau(x),\tau(y)},\quad\tau(x)=1,\dots,P\,,
\eeq
where $K$ represents the coupling to an auxiliary $P$-state Potts variable $\tau(x)$.
In this way a renormalization group analysis in $J$ and $K$ can be performed,
followed by the limit $P\to 1$ to get rid of the auxiliary degrees of freedom.
In two dimensions one finds two non-trivial fixed points, a repulsive one at $K=J_c$ and an attractive one
at $K=K^*>J_c$, both for $J=J_c$, see Fig.~\ref{PD}. Hence the critical behavior of the spin clusters, which correspond to $p=1$, i.e. $K=+\infty$, is controlled by the fixed point with $K=K^*$. It also 
follows from the analysis of \cite{CP} that at the fixed point with $K=J_c$ the scaling dimension $X$ in (\ref{P2}) coincides with that of the Potts spin field, as it should for the FK clusters \cite{FK}. This scaling dimension takes the value $X_{FK}=X_{1/2,0}$ \cite{Nienhuis} within the Kac parameterization
\EQ
X_{m,n}=\frac{[(t+1)m-tn]^2-1}{2t(t+1)}\,,
\label{kac}
\EN
where $t$ is related to $Q$ as 
\EQ
\sqrt{Q}=2\sin\frac{\pi(t-1)}{2(t+1)}\,.
\label{tq}
\EN

Since the only physical degrees of freedom are those of the Hamiltonian (\ref{hamiltonian}), both fixed points have the $Q$-state Potts central charge \cite{DF,paraf}
\EQ 
c=1-\frac{6}{t(t+1)}\,.
\label{c}
\EN

Concerning the fractal dimension of spin clusters, a distinction has to be made between $Q=2$ and $Q\neq 2$.
For $Q=2$, \eqref{hamDP} is  the ordinary Hamiltonian of the dilute $P$-state
Potts model\footnote{The binary Ising spin variable plays the role of the vacancy.},
and this implies that the spin cluster fixed point is identified with the tricritical point
of the $P\to 1$ Potts model \cite{CK}; this fixes $X_S|_{Q=2}=5/96$, as first pointed out in \cite{SV}.
Such a neat mapping is not  available at $Q\neq 2$. A conjecture for $X_S$, however, was made in \cite{Vanderzande}
arguing triviality ($X_S=0$) at $Q=1$ and coalescence of FK and spin cluster fixed points at $Q=4$. Then, assuming the form
(\ref{kac}) with $t$-independent indices and using the values at $Q=1,2,4$ (i.e. $t=2,3,\infty$), one obtains
$X_S=X_{0,1/2}$ \cite{Vanderzande}. The corresponding value $X_S|_{Q=3}=7/80$ agrees with numerical determinations
\cite{Vanderzande,DBN,DJS,DJS2}.

The scaling dimension $X_{0,1/2}$ is that of the spin field along the tricritical branch of the Potts model \cite{Nienhuis}.
The critical and tricritical branches meet at $Q=4$ and can be considered as continuation of each other \cite{Wu}.
Indeed, within the parameterization $Q=2+2\cos(g\pi/2)$, $g\in(2,4)$ corresponds to the critical branch (where $t=g/(4-g)$),
$g\in(4,6)$ corresponds to the tricritical branch (where $t=4/(g-4)$), and the scaling dimensions on both branches read
$1-g/8-3/2g$ for the spin field, $6/g-1$ for the energy field, and so on \cite{Nienhuis}.
From this starting point, a picture in which spin clusters are seen as continuation of FK clusters
to the tricritical branch\footnote{It is important to stress that, as far as percolative properties are concerned, the point to be selected on the tricritical branch is that with the central charge prescribed by (\ref{tq}), (\ref{c}).} has been further discussed in the literature (see e.g. \cite{DBN2,JS}). The picture has a counterpart in the 
framework of Schramm-Loewner evolution (SLE), in which boundaries of critical clusters are associated to stochastic
curves with conformal invariant measure. Using this approach, the fractal dimension of Ising spin cluster boundaries
has been rigorously derived in \cite{Smirnov}, while a numerical investigation of the case $Q=3$ is performed in \cite{GC}.
A related duality relation for fractal dimensions of cluster boundaries was obtained in \cite{Duplantier}.

The argument for (\ref{Rfk}) in \cite{DV} exploits duality to relate the two- and three-point functions of the
spin field to those of disorder fields $\mu_{ab}$ associated to boundaries between clusters of colors $a$ and $b$.
It is then observed that, as long as two- and three-point functions are concerned (see also \cite{duality}), these fields
can be replaced by a doublet $\mu$, $\bar{\mu}$, and that those correlators can be rewritten as correlators of a single
field $\Phi=(\mu+\bar{\mu})/\sqrt{2}$. The correlators of $\Phi$ can then be evaluated within the theory with non-degenerate
spectrum of \cite{AlZ} and produce\footnote{We do not reproduce here the structure constant $C(Y)$. As explained above
it is a particular case of the results of \cite{AlZ}, reproduced in \cite{DV,ZSK} with the notations of this paper.} $C(X_{FK})$, while the $\sqrt{2}$ remembers that $\Phi$ originates from a doublet and is the only remnant of color symmetry.
It is not clear how to extend this type of considerations to spin clusters, despite the fact that the symmetry to deal with is always $S_Q$. The Ising case, the only one for which the relation with the tricritical branch is clear, is discussed in the final section.

\begin{figure}[t]
\begin{center}
\epsfxsize=450pt\epsfysize=350pt{\epsffile{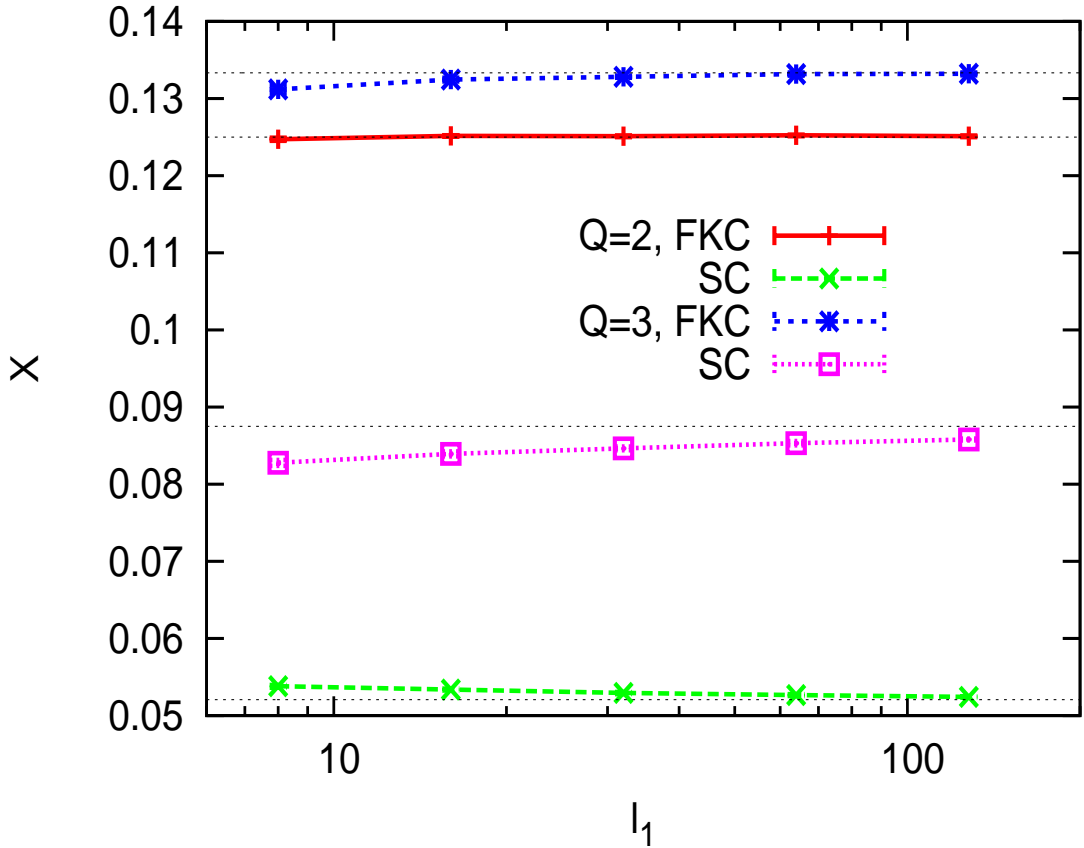}}
\end{center}
\caption{From top to bottom the dotted lines correspond to $X_{FK}|_{Q=3}=2/15$, $X_{FK}|_{Q=2}=1/8$, $X_S|_{Q=3}=7/80$, $X_S|_{Q=2}=5/96$. The numerical data obtained from (\ref{max}) are plotted versus the size parameter $l_1$ defined in the text.
}
\label{PQ_mag}
\end{figure}
\section{Numerical results}
We now present our simulated results for various values of $Q$ for the spin cluster connectivity constant $R_S$ 
defined as in (\ref{P3}). Our simulations were done on square lattices of linear size $L$ and 
with periodic boundary condition in both directions. We employ the Wolff cluster algorithm to equilibrate the system \cite{Wo}. For non integer values of $Q$, 
we employ the Chayes-Machta algorithm \cite{CM,DGMOPS} which is an extension of the Swendsen-Wang algorithm. 
For each value of $Q$, we perform measurements over $10^6$ independent configurations  for $ 8 \leq L \leq 2048$.
For $Q \leq 3$, we also simulate
$2\times 10^5$ independent configurations  for $L=4096$ and $10^5$ for $L=8192$. 

The computation of the spin cluster connectivity constant $R_S$ is in principle straightforwardly done by 
computing the three-point connectivity. This is done by considering, for each point $(i,j)$ 
on the lattice, the correlation  $P_3^{S}((i,j),(i + \Delta,j), (i,j+\Delta))$ as a function of $\Delta$. 
This is the method that we employed in our previous study for the FK cluster connectivity constant $R_{FK}$ \cite{PSVD}. 
In practice, we discovered that while this method works well for small values of $Q$ (up to $\simeq 2$), it is not any more the case as $Q$ becomes larger due to 
strong finite size corrections. The value of $R_S$ is not converging even for the largest sizes that we simulate.  
We will first start explaining why by considering measurements of simple quantities like the behavior of the fractal dimension $2-X_S$ of the spin clusters. 
If $\langle v_{max}\rangle$ denotes the average over spin configurations of the size of the largest cluster, one expects 
\beq
\langle v_{max}\rangle\sim L^{2-X}
\label{max}
\eeq
at the critical point on a lattice of linear size $L$. As seen in the previous section, in terms of (\ref{kac}) and (\ref{tq}), the scaling dimension $X$ is 
\EQ
X_S=X_{0,1/2}=\frac{t^2-4}{8t(t+1)}
\label{xs}
\EN
for spin clusters (only conjectured for $Q\neq 2$), and 
\EQ
X_{FK}=X_{1/2,0}=\frac{t^2+2t-3}{8t(t+1)}
\label{xfk}
\EN
for FK clusters. In Fig.~\ref{PQ_mag} we show the values of $X_{FK}$ and $X_S$  measured for $Q=2$ and $Q=3$ using (\ref{max}).
The exponents are obtained by a 5-point fit with data for lattice sizes $2^n l_1$, $n=0,1,\ldots,4$; $l_1$ is shown
on the horizontal axis. As can be seen, the convergence towards the theoretical values is excellent for the FK clusters and quite good for the Ising
spin clusters.%
\begin{figure}[t]
\begin{center}
\epsfxsize=450pt\epsfysize=350pt{\epsffile{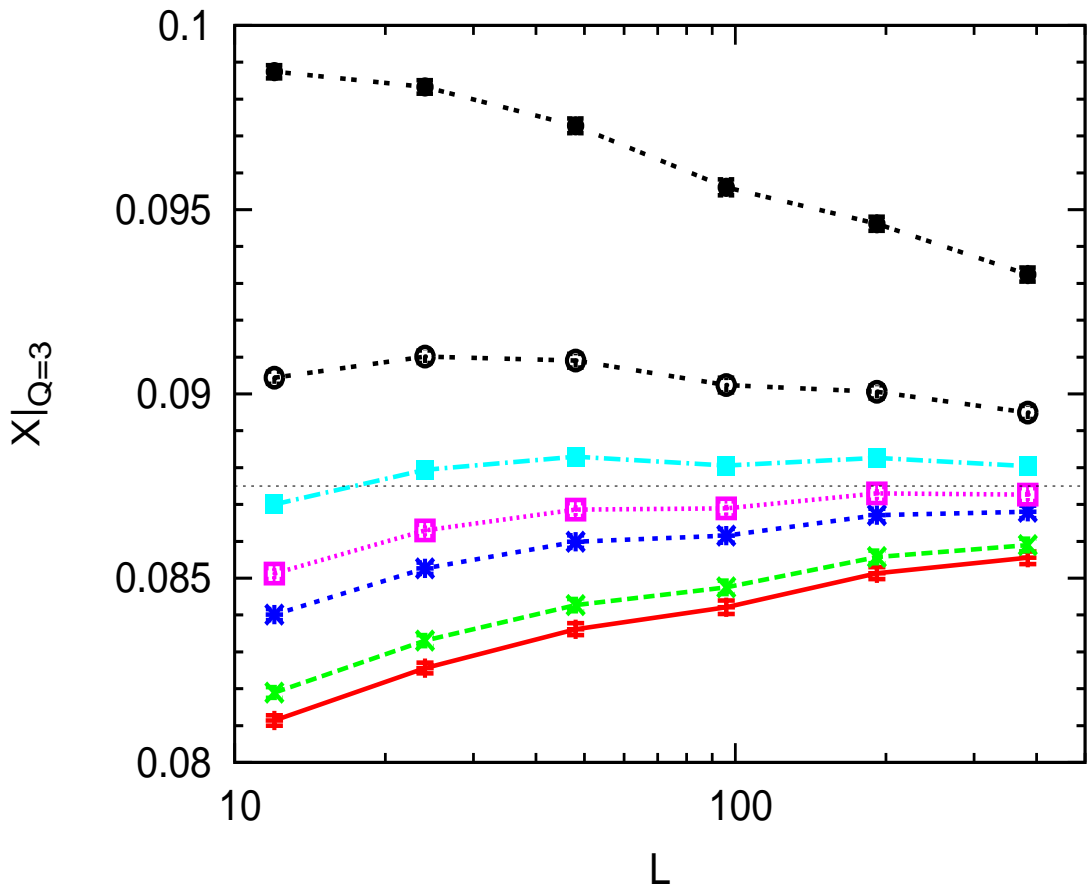}}
\end{center}
\caption{The exponent $X|_{Q=3}$ measured through (\ref{max}) for generalized clusters obtained connecting nearest neighbors with the same color with probability $1-e^{-K}$. The size $L$ is on the horizontal axis. The different curves correspond, from bottom to top, to $K=\infty, 3 J_c, 2 J_c, 1.8 J_c, 1.6 J_c, 1.4 J_c, 1.2 J_c$. The dotted line is the conjectured value $X_S|_{Q=3}=7/80$. 
}
\label{PQ_mag2}
\end{figure}
The convergence is very slow, instead, for the spin clusters at $Q=3$. 
Even for the largest sizes, with data from $l_1=128$ up to $2^4l_1=2048$, we observe a deviation of $2 \%$ between our data and the expected value. 
\begin{figure}[t]
\begin{center}
\epsfxsize=450pt\epsfysize=350pt{\epsffile{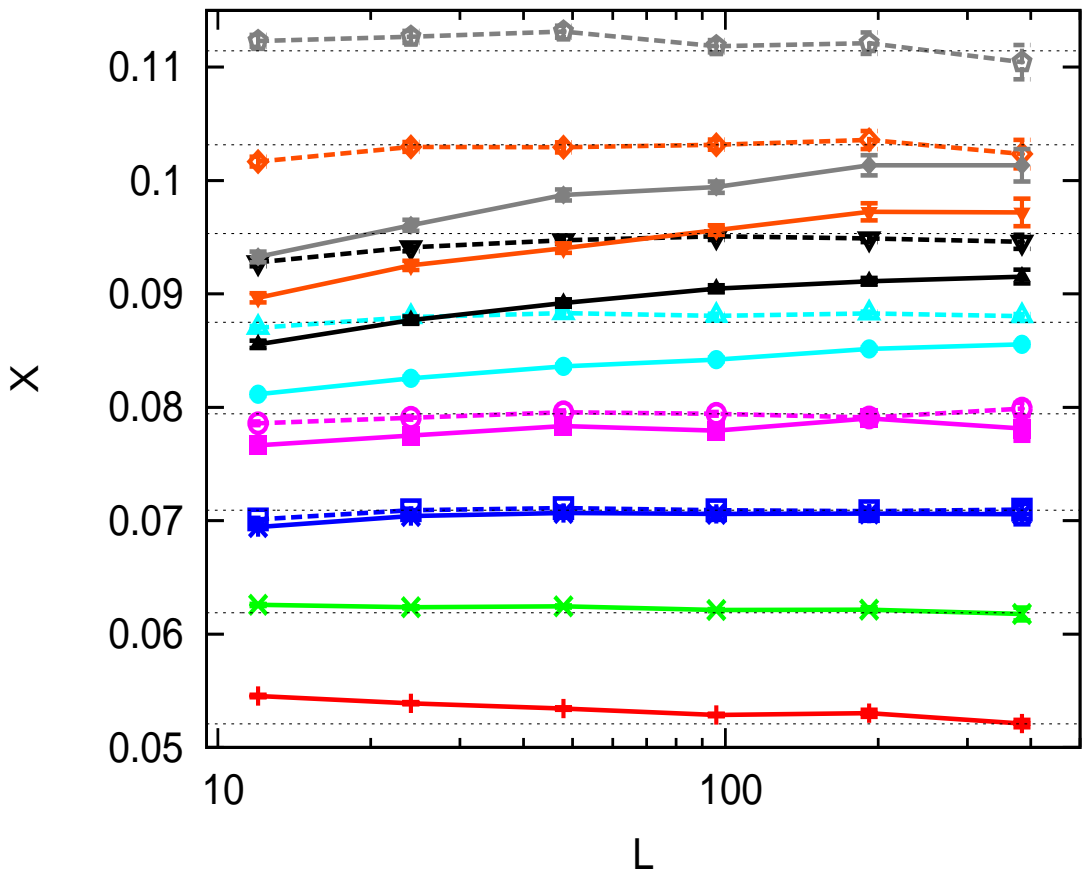}}
\end{center}
\caption{The exponent $X$ measured through (\ref{max}) for generalized clusters obtained connecting nearest neighbors with the same color with probability $1-e^{-K}$. The continuous lines connect the points for the spin clusters ($K=+\infty$), while the dashed lines correspond to the exponent for the values $K^*$ indicated in Tab.~\ref{Table1}. 
The size $L$ is on the horizontal axis. The different colors correspond, from bottom to top, to $Q= 2.0, 2.25, 2.5, 2.75, 3.0, 3.25, 3.5, 3.75$. For each of these values of $Q$ we also show as a dotted line the conjectured value $X_S$.
}
\label{PQ_mag3}
\end{figure}

Measurements done on two- and three-point connectivities show a similar convergence pattern,
as we have checked in detail. In particular, as a consequence of the quite poor convergence for $Q=3$ spin clusters,
we observed that the ratio $R_S|_{Q=3}$ built from two- and three-point spin cluster connectivities does not converge
(see the right lower part of Fig.~\ref{PQ3}). This suggested to us that, within the renormalization group picture
discussed in the previous section, renormalization from $K=+\infty$ towards the fixed point value $K^*$ 
ruling the universal properties of spin clusters is significantly slower for $Q=3$ than for $Q=2$.
We then simulated the generalized clusters for $Q=3$, with the aim of determining $K^*$,
i.e. we looked for the value of $K$ for which the exponent converges faster.
In order to better observe the crossover towards the fixed point $K^*$ we plot in Fig.~\ref{PQ_mag2}
the exponent $X|_{Q=3}$
obtained by a two-point fit for linear size $L$ and $2 L$, as a function of lattice size and for different values of $K$.
To the same purpose, we also increased the statistics and collected 
$10^7$ independent configurations, which is ten times more than for the other measurements. We recall that $J_c=\log{(1+\sqrt{Q})}$ on the square lattice. The figure shows that there are very few corrections in the range $K = 1.6 J_c - 1.8 J_c$. Hence we will 
take the value $K^*\simeq 1.7 J_c$ in the following. Moreover, we observe a clear crossover towards this point, confirming that it is an attractive fixed point. The analysis also clearly supports the conjectured value $X_S|_{Q=3}=7/80$.
\begin{figure}[t]
\begin{center}
\epsfxsize=450pt\epsfysize=350pt{\epsffile{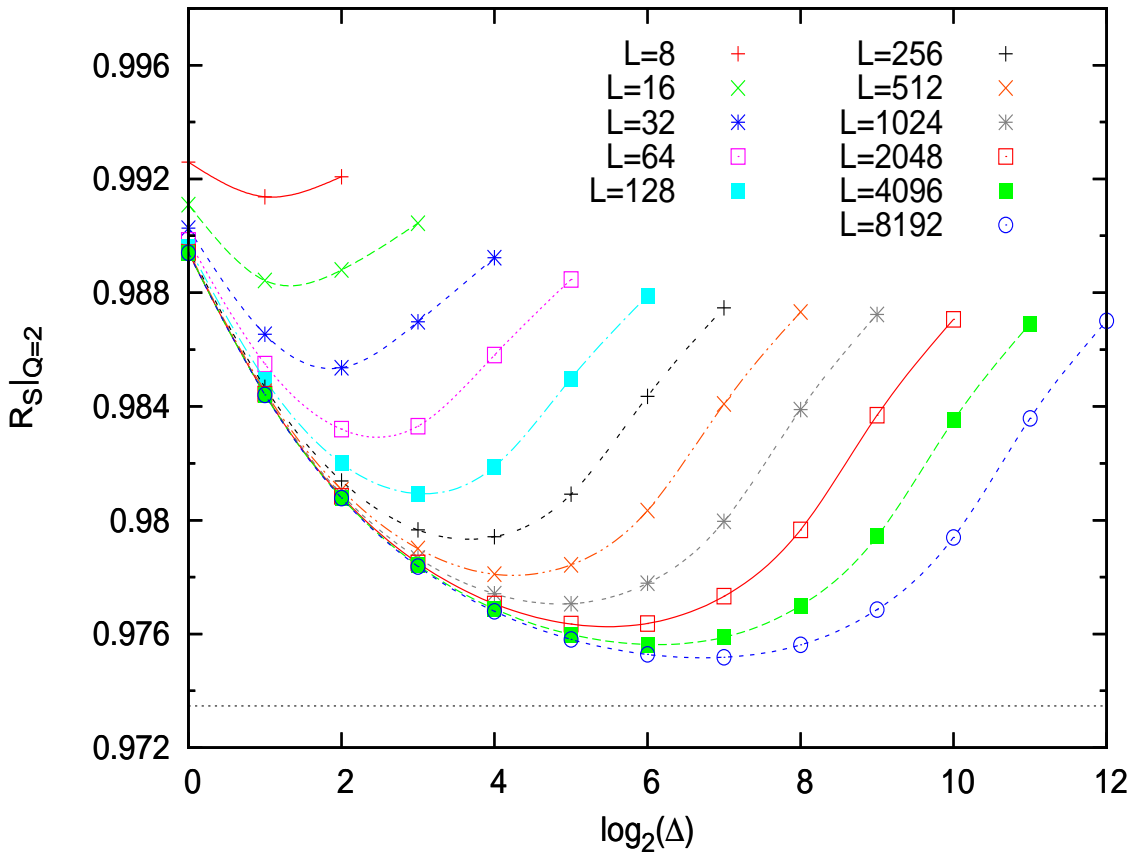}}
\end{center}
\caption{
Values of the connectivity ratio $R_S|_{Q=2}$ as a function of $\log_2(\Delta)$. The different curves correspond to interpolation between 
the points obtained for different linear sizes $L$ as shown in the caption.
The dotted line corresponds to $C(X_S)|_{Q=2}=0.973474$. 
}
\label{SC2}
\end{figure}

The same type of the measurement of the cluster fractal dimension $2-X$ was also done for $Q=1.25, 1.5, 1.75, 2.25, 2.5, 2.75, 3.25, 3.5, 3.75, 4$. For $Q < 2.5$, we observe that $X$ converges faster for $K = +\infty$, as it was the case for $Q=2$; in this range of $Q$ we obtained in this way a good agreement with the prediction from the spin field on the tricritical branch  given by (\ref{xs}).
For $Q \geq 2.5$, we found a finite value of $K^*$ for which the measured $X$ converges faster towards the expected value, as it was the case for $Q=3$; in this range of $Q$ we will then use in the following the generalized clusters with $K=K^*$. Our  measured  values of $K^*/J_c$ are reported in Tab.~\ref{Table1}.
\begin{figure}
\begin{center}
\epsfxsize=220pt\epsfysize=180pt{\epsffile{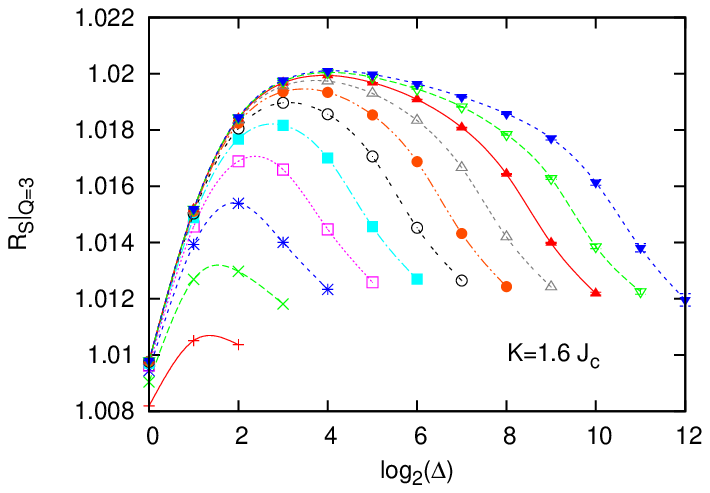}}
\epsfxsize=220pt\epsfysize=180pt{\epsffile{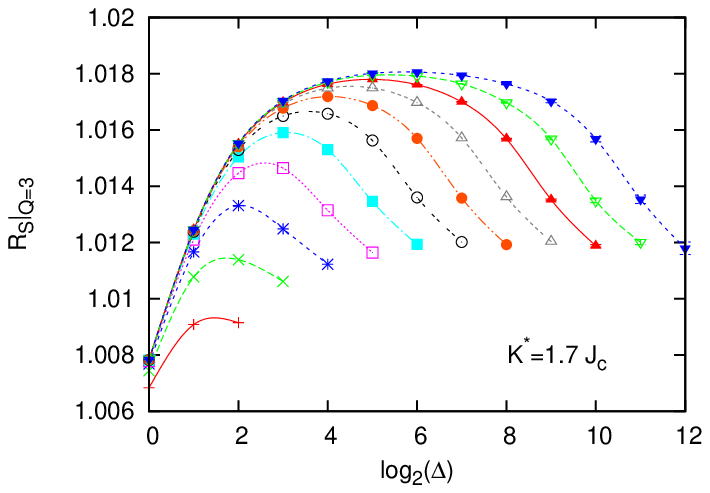}}\\
\epsfxsize=220pt\epsfysize=180pt{\epsffile{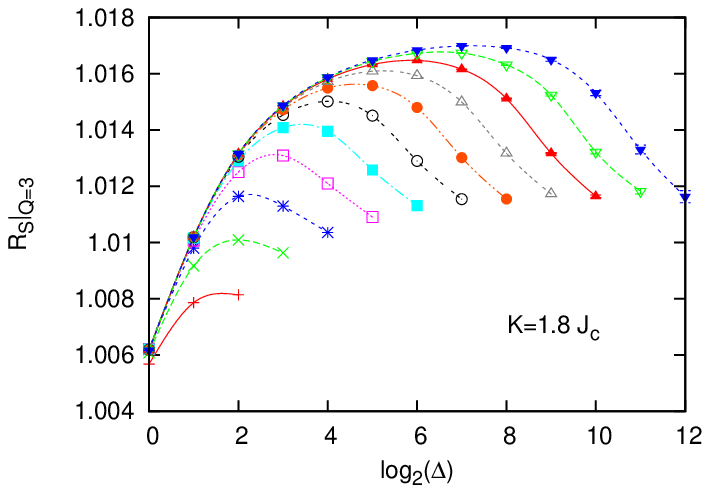}}
\epsfxsize=220pt\epsfysize=180pt{\epsffile{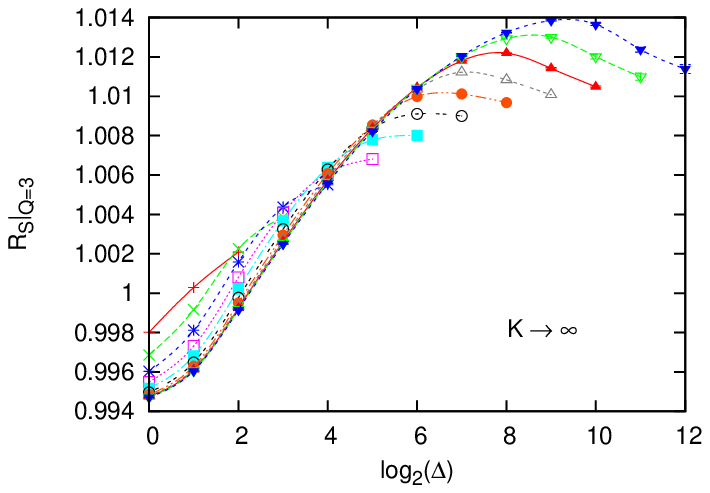}}
\end{center}
\caption{
Values of the connectivity ratio $R|_{Q=3}$ vs $\log_2(\Delta)$ for generalized clusters with different value of the parameter $K$. The symbols for the different sizes $L$ are the same as in Fig.~\ref{SC2}.
}
\label{PQ3}
\end{figure}
%
%
\begin{table}
\caption{Determinations of the fixed point $K^*$ for different values of $Q$. }
\begin{center}
\begin{tabular}{ | l ||  c | c | c | c |  c |  c | c | }
\hline
  $Q$ & 2.5 & 2.75 & 3.0 & 3.25 & 3.5 & 3.75 & 4 \\ \hline 
  $K^*/J_c$ & 3.0 (5)  & 2.1 (2)   & 1.7 (1) & 1.5 (1)  & 1.3 (1)  &  1.15 (5) &  1.0  \\
  \hline
\end{tabular}
\label{Table1}
\end{center}
\end{table}
The results for $X$ measured through (\ref{max}) for several values of $Q$ in the range $2 \leq Q < 4$ are shown in Fig.~\ref{PQ_mag3} as a function of lattice size. The data points connected by continuous lines are obtained at $K=+\infty$ by a two-point fit for linear size $L$ and $2L$ (and the line is just a interpolation). The dashed lines connect instead the points corresponding to the measurement with the values $K^*$ of Tab.~\ref{Table1}. The dotted lines give the conjectured value of $X_S$ for each of the considered values of $Q$. Note 
that the sign of the corrections for the effective exponent changes between $Q=2.25$ and $Q=2.5$. This change, as we increase $Q$, corresponds to the appearance of an attractive fixed point for a finite value of $K^*$. This is in qualitative agreement with the results of Deng et al. \cite{DBN2} (see in particular their Fig.~2).

In Fig.~\ref{PQ_mag3},  we did not show the results for $Q < 2$ and $Q=4$. 
For $Q < 2$, convergence towards the conjectured value $X_S$ is excellent, as we have also checked. For $Q=4$, the convergence is not very good already for the exponent $X_{FK}$ associated to the FK clusters, which is not a surprise since we expect logarithmic corrections for this value of $Q$.
Still, even with these logarithmic corrections, it is possible to see that for the generalized clusters the exponent converges faster for values of $K$ very close to $J_c$, i.e. not exceeding $1.1 J_c$, consistently with coalescence of the fixed points $K=J_c$ and $K=K^*$ as $Q\to 4$. 

We now go to the main part of our work, the computation of the spin cluster connectivity constant $R_S$. 
From the previous discussion on the behavior of the scaling dimension $X_S$, we know that for small values of $Q$ (up to $Q=2.25$) a measurement of
the  three-point connectivity directly for the spin clusters ($K=+\infty$) must allow us to obtain $R_S$. 
Our numerical results for $R_S$ at $Q=2$ are shown in Fig.~\ref{SC2}, as a function of $\log_2\Delta$ and for increasing values of the linear size $L$. As $L$ increases, a convergence towards the constant value corresponding to the continuum theory on the infinite plane is expected to emerge for values of $\Delta$ much larger than lattice spacing and sufficiently smaller than $L$.
Such a pattern is indeed visible in Fig. \ref{SC2}. More quantitatively, it is apparent from Fig. \ref{SC2} that the curves have a minimum for $\Delta = \sqrt{L/2}$  which lies in the bulk asymptotic region $1 \ll \Delta \ll L$. Then we can read the asymptotic value $R_0$ from a fit of the form~:
\beq
\label{fitR}
R_S(L,\Delta=\sqrt{L/2}) = R_0 + R_1 L^{-\omega}\;.
\eeq
We obtain (keeping only the values for which $\sqrt{L/2}$ is a power of $2$) 
$R_0|_{Q=2}= 0.9737 (2)$, very close to the value $C(X_S|_{Q=2})=0.973474..$\,.
A similar analysis has also been performed for other values of $Q \leq 2.25$. The results for $R_0$ are reported in Tab.~\ref{Table2}. 

\begin{figure}[t]
\begin{center}
\epsfxsize=450pt\epsfysize=350pt{\epsffile{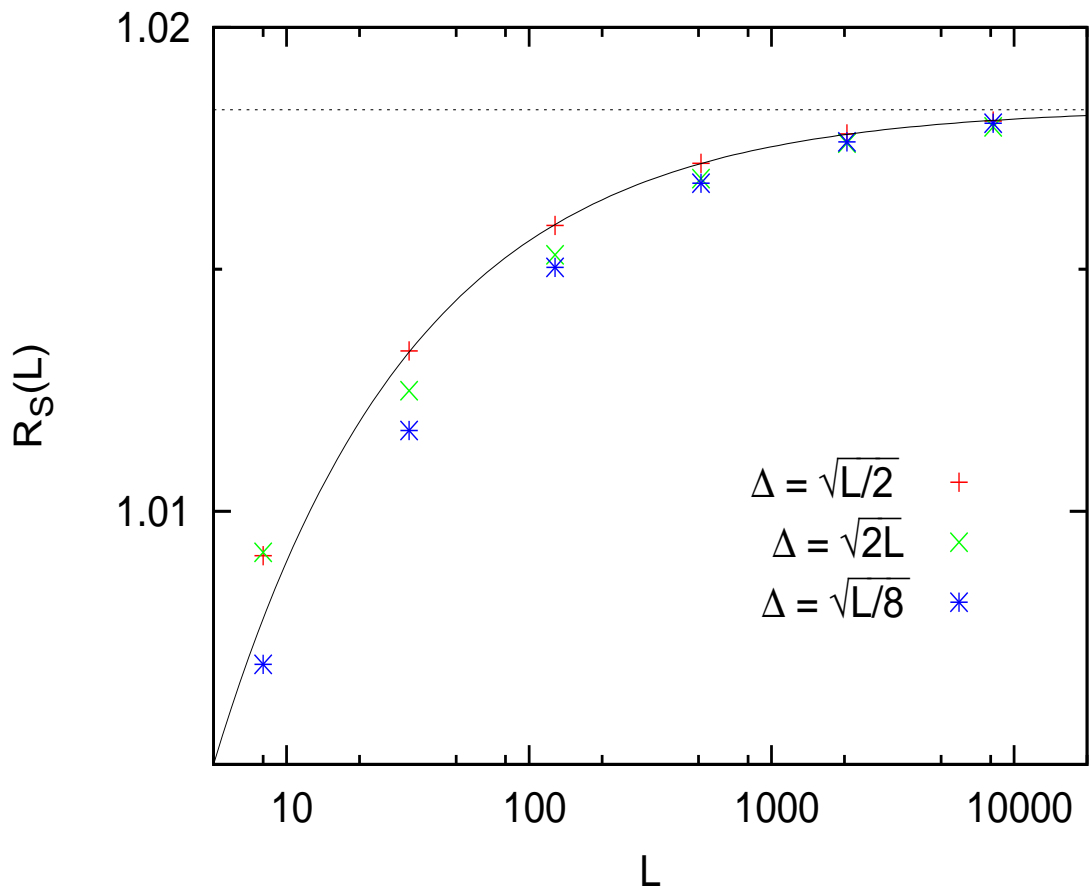}}
\end{center}
\caption{
Measured values of $R_S(L,\Delta)$ vs. $L$ for $\Delta = \sqrt{L/2}, \sqrt{L/8}$ and $\sqrt{2L}$. The continuous line is a plot to the form Eq.(\ref{fitR}) and the dotted line corresponds to $R_0|_{Q=3} = 1.0183$. 
}
\label{PQ3b}
\end{figure}

Next we turn to the analysis of the range $Q\geq 2.5$, starting with the case $Q=3$. As anticipated above, at $Q=3$ the ratio $R_S$, if measured for the ordinary spin clusters, does not converge, even for the largest sizes considered, as can be seen in the right lower part of Fig.~\ref{PQ3}. A direct inspection of the two- and three-point connectivities shows that corrections to scaling are extremely strong, in particular for the three-point case. The expectation is that these corrections are minimized if we consider the clusters corresponding to the attractive fixed point $K^*$. The results shown in Fig.~\ref{PQ3} for $K=1.6J_c,1.7J_c,1.8J_c$ indeed confirm $1.7J_c$ as the $Q=3$ fixed point value, for which a fit of the form (\ref{fitR}) converges nicely to $R_0|_{Q=3} = 1.0183 (5)$. We show in Fig.~\ref{PQ3b} the data for the measured values of $R_S(L,\Delta=\sqrt{L/2})$ as a function of $L$ and 
a plot to the form Eq.(\ref{fitR}) with $32 \leq L \leq 8192$. We also show in this figure the measured values of $R_S(L,\Delta=\sqrt{L/8})$ and $R_S(L,\Delta=\sqrt{2L})$. A fit of these points to the form Eq.(\ref{fitR}) leads to the same value of $R_0|_{Q=3}$ if we take in account the error bars. 
%
For the other values in the range $2.5 < Q < 4$, the same type of analysis produces the results reported in Tab.~\ref{Table2}. For $Q=2.5$ we observed that the spin clusters ($K=+\infty$) and the generalized clusters at $K^* \simeq 3.0 J_c$ produced essentially the same result for the connectivity ratio. For $Q > 3$, we have data only up to size $L=2048$. A fit to the form 
(\ref{fitR}) is difficult to be done without data for larger sizes. So for these values of $Q$, we modified our analysis. For each size $64 \leq L \leq 2048$, we determined the maximum value 
$R_S(L, \Delta)$ as a function of $\Delta$ with a parabolic fit. Then we employ these maximum in place of $R_S(L,\Delta=\sqrt{L/2})$ in a fit to the form (\ref{fitR}). We checked on 
our data for smaller values of $Q$ that this method produces the same results as with a fit with $R_S(L,\Delta=\sqrt{L/2})$ but with slightly larger error bars. 

\begin{figure}[t]
\begin{center}
\epsfxsize=450pt\epsfysize=350pt{\epsffile{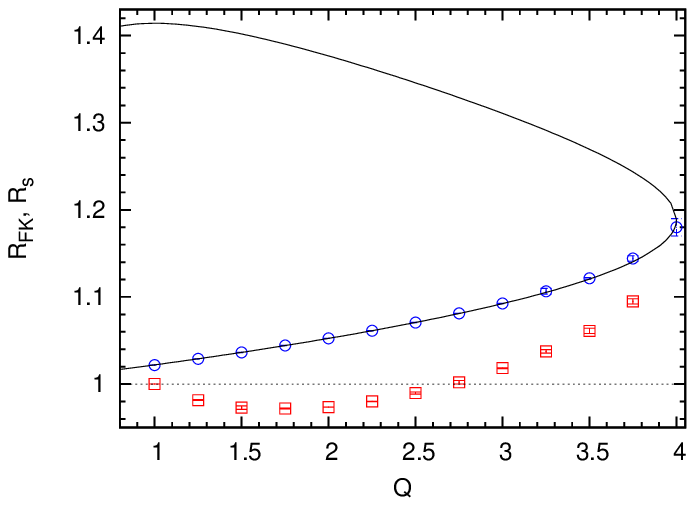}}
\end{center}
\caption{
Data points for $R_{FK}$ (blue circles) and for $R_S$ (red squares). The curve is $\sqrt{2}C(X)$ with $X=X_{FK}$ (lower branch) and $X=X_S$ (upper branch).
}
\label{R}
\end{figure}

Concerning $Q=4$, the result for the connectivity ratio  $R_{FK} = 1.18 (1)$ is in fair agreement  with the prediction $\sqrt{2}C(X_{FK})=1.1892..$~. Note that the present numerical value is slightly different, but compatible, with the one reported in \cite{PSVD}.
The difference is due to the new analysis that we just explained above for the case with data only up to $L=2048$. 
All our results are summarized in Tab.~\ref{Table2} and Fig.~\ref{R}
\begin{table}
\caption{Numerical results for $R_S$ and $R_{FK}$ as a function of $Q$}
\begin{center}
\begin{tabular}{ | l ||  c | c | c | c |  c |  c |  c |}
\hline
  $Q$          & 1.0           & 1.25           &   1.5          &   1.75        & 2.0             & 2.25          & 2.5                    \\ \hline 
  $R_{s} $   & 1.0           & 0.9815 (5)  & 0.973 (2)  & 0.9720 (5) & 0.9735 (2) & 0.9800 (3) & 0.9896 (12)    \\
  $R_{FK}$ & 1.0218 (2)& 1.0290 (2)  & 1.0364 (2)& 1.0442 (2) & 1.0524 (2) & 1.0613 (2) & 1.0706 (2) \\
  \hline
 $Q$    &       & 2.75         & 3.0            & 3.25        & 3.5           & 3.75        & 4.0         \\ \hline 
 $R_{s} $  &     & 1.002 (2) & 1.0183 (5) & 1.0376 (20) & 1.061 (3) & 1.093 (3) & 1.18 (1)  \\
 $R_{FK}$  &  &  1.0811 (2)&1.0925 (2) & 1.1065 (33) & 1.1215  (10) & 1.1446 (33) & 1.18 (1) \\
\hline
\end{tabular}
\label{Table2}
\end{center}
\end{table}

\section{Discussion}
In this paper we studied numerically the spin clusters of the critical $Q$-state Potts model with Hamiltonian (\ref{hamiltonian}) at regular intervals of width $0.25$ in the range $1<Q\leq 4$. More generally, we studied the clusters obtained connecting nearest neighboring sites with the same color with probability $1-e^{-K}$, the spin clusters corresponding to $K=+\infty$. The numerical results confirm the presence of a repulsive fixed point at $K=J_c$, corresponding to FK clusters, and of an attractive fixed point at $K=K^*>J_c$, ruling the universal properties of spin clusters. The results are consistent with coalescence of the two fixed point as $Q\to 4$, meaning that spin and FK clusters belong to the same universality class in this limit. Working at $K^*$ when needed to improve the convergence, we obtained accurate estimates of the fractal dimension $2-X_S$ of spin clusters, in very good agreement with the conjectured value (\ref{xs}), (\ref{tq}). As recalled in section~
 2 this value corresponds to a 
continuation to the tricritical Potts branch of the scaling dimension (\ref{xfk}) associated to FK clusters.

We then measured the universal three-point connectivity constant $R_S$ of spin clusters, defined as in (\ref{P3}), obtaining
again quite accurate results. As clearly apparent from Fig.~\ref{R}, for this observable the naive continuation to the
tricritical branch, amounting to replacing $X_{FK}$ with $X_S$ in the result (\ref{Rfk}) for FK clusters, fails. This is not
really surprising once we have verified that a critical line for spin clusters exists for real values of $Q$ down to $Q=1$.
Indeed, as $Q\to 1$ the whole lattice forms a single spin cluster and all connectivities tend to 1, meaning that also $R_S$
tends to 1. In field theoretical terms the constance of the connectivities means that the underlying field is the identity ($X_S=0$), consistently with the fact that the unique cluster has trivial fractal dimension 2. It can be checked that the structure constant $C(0)$ gives 1, as it should since it is computed within a theory with the identity as the only field with
 vanishing dimension. Hence, $R_S$ 
has to tend to $C(X_S)$ as $Q\to 1$, and not to $\sqrt{2}\,C(X_S)$.

The fact that $R_{FK}$ and $R_S$ are not continuation of each other indicates that critical FK clusters and critical spin clusters are described by conformal field theories with different realizations of the color symmetry $S_Q$ of the Potts model. The sensitivity of the three-point connectivity to the realization of the symmetry is apparent in the result (\ref{Rfk}), which factorizes into a piece (the constant $\sqrt{2}$) accounting for color symmetry, and a structure constant computed within a colorless CFT. The results of this paper show that, if such a factorization makes sense also for spin clusters, it is not as simple, since the color factor will be $Q$-dependent. 

For the Ising case $Q=2$, the numerical result $R_S=0.9735(2)$ points towards exact coincidence with $C(X_S)=0.97347..$\,. Theoretically, such a coincidence is plausible. Indeed, for $S_2=Z_2$ symmetry the spectrum of scaling dimensions is non-degenerate, and this is the case for which the structure constants of \cite{AlZ} are derived. Moreover, the spin three-point function on the tricritical branch is computed not at the Ising tricritical point with central charge $7/10$, where it would vanish by symmetry as in the case of FK clusters, but at the point with the central charge $1/2$ of the critical Ising model, where it has no reason to vanish and to require additional arguments. Such a different role of the symmetry for spin clusters extends to the other values of $Q$ in a way whose implementation in the conformal theory remains to be understood.

\noindent
{\bf ACKNOWLEDGMENTS:}
We thank J. Jacobsen, R. Vasseur and in particular Y. Ikhlef for many, enlightening discussions. R.S.  acknowledges support by ANR grant
2011-BS04-013-01 WALKMAT.


\end{document}